\documentclass[english,twocolumn,english,aps,prx]{revtex4-1}
\usepackage[LGR,T1]{fontenc}
\usepackage[latin9]{inputenc}
\usepackage{amsmath}
\usepackage{amssymb}
\usepackage{graphicx}

\makeatletter

\DeclareRobustCommand{\greektext}{%
  \fontencoding{LGR}\selectfont\def\encodingdefault{LGR}}
\DeclareRobustCommand{\textgreek}[1]{\leavevmode{\greektext #1}}
\ProvideTextCommand{\~}{LGR}[1]{\char126#1}

\usepackage{mathrsfs}

\makeatother

\usepackage{babel}
\begin{document}

\title{Dark State Polarizing a Nuclear Spin in the Vicinity of a Nitrogen-Vacancy
Center}

\author{Yang-Yang Wang$^{1,3}$, Jing Qiu$^{1,3}$, Ying-Qi Chu$^{1,3}$,
Mei Zhang$^{1}$, Jianming Cai$^{2}$, Qing Ai$^{1,*}$, and Fu-Guo
Deng$^{1}$}

\address{$^{1}$Department of Physics, Applied Optics Beijing Area Major Laboratory,
Beijing Normal University, Beijing 100875, China}

\address{$^{2}$School of Physics and Center for Quantum Optical Science,
Huazhong University of Science and Technology, Wuhan 430074, China}

\address{$^{3}$These authors contribute equally to the work.}
\email{aiqing@bnu.edu.cn}

\date{\today}
\begin{abstract}
The nuclear spin in the vicinity of a nitrogen-vacancy (NV) center possesses of long coherence time and convenient manipulation assisted by the strong hyperfine interaction with the NV center. It is suggested for the subsequent quantum information storage and processing after appropriate initialization. However, current experimental schemes are either sensitive to the inclination and magnitude of the magnetic field or require thousands of repetitions to achieve successful realization. Here, we propose polarizing a $^{13}$C nuclear spin in the
vicinity of an NV center via a dark state. We demonstrate theoretically that it is robust to polarize various nuclear spins with different hyperfine couplings and noise strengths.

\end{abstract}
\maketitle

\section{Introduction}

Benefiting from quantum entanglement \cite{Trifunovic2013,Dobrovitski2013},
quantum information processing \cite{Nemoto2014} can effectively
speed up computation and ensure security of information \cite{Nielsen2000}.
As the basic element, quantum-bit (qubit) lies at the heart of quantum
information processing \cite{Nielsen2000,Fuchs2009}. Solid-state
qubits are a promising candidate because they might well explore the
well-developed technology of semiconductor industry \cite{Hanson2008,Cai2013}.
Remarkably, the nitrogen-vacancy (NV) center in diamond has been recognized
as an intriguing choice since it is of easy accessibility and long
coherence time at room temperature \cite{Doherty2016,Doi2014,Barclay2011,Anjou2016,Toyli2012},
as can be measured by Landau-Zener-St\"{u}ckelberg interferometry
\cite{Huang2011-2}. To date, various applications including quantum
information processing and quantum metrology have been successfully
realized in the NV centers \cite{Buckley2010,Mamin2013,Zhang14}.
For example, different versions of transitionless driving algorithms
have been fully utilized to accelerate quantum control in the NV centers
\cite{Zhou2017,Baksic2017,Zhang2013,Song2016}. Besides, the NV centers have
been explored to detect internal dynamics of clusters of nuclear spins
by dynamical decoupling \cite{Zhao2011-1,vanderSar2012} with sensitivity
further improved by Fluoquet spectroscopy and coupling to collective
modes \cite{Lang2015,Ajoy2015,Wolf2015}. The NV center has also been
proposed to detect the radical-pair chemical reaction in biology \cite{Liu2017}.
Due to the quantum nature of surrounding nuclear spin bath, the anomalous
decoherence effect of the NV center has been theoretically predicted
\cite{Zhao2011} and experimentally verified \cite{Huang2011}.

Apart from the electron spin of the NV center, the nuclear spins in
the vicinity of an NV center is of broad interest to the community.
Due to much longer coherence time, nuclear spins are more frequently
used in the quantum information storage and processing \cite{Childress2006,Dutt2007,Jacques2009,Jiang2009,Neumann2010,Maurel2012,Shim2013,Zu2014,Zhang2015}.
However, it is difficult to initialize and control the nuclear spins
because of their small magnetic moments. Utilizing an ancillary electronic
spin to couple with the nuclear spin by the hyperfine interaction
may effectively overcome this limitation.

To our best knowledge, there are three kinds of experimental schemes
which have been successfully demonstrated to initialize the nuclear
spins around the NV centers. A straightforward approach is to repeatedly
perform projective measurements until the desired state is observed
\cite{Neumann2010,Jiang2009}. An alternative is to bring the excited
(ground) state close to the level-anticrossing point by applying a
specific static magnetic field \cite{Jacques2009,Wang2013}. In the
last but widely-used approach \cite{Childress2006,Dutt2007,Shim2013,Neumann2008,Taminiau2014},
the electron spin is first initialized, and then its polarization
is coherently swapped to the nuclear spin. After tens of repetitions
of the above process, nearly-complete polarizations of both electronic
and nuclear spins are achieved. Here, we remark that in each repetition
the swapping of polarization between the electron and nuclear spins
is essentially a quantum-state transfer process.

For any state transfer process, the fidelity is inevitably influenced
by the noise due to coupling to the bath.  On the other hand,
we notice that the dark state has been extensively
applied to coherently transfer energy in photosynthetic light harvesting
\cite{Dong2012} and perfectly transfer state in optomechanical systems
\cite{Wang2012,Khanaliloo2015}. The coherent coupling between a surface
acoustic wave and an NV center has been experimentally realized via
the dark state recently \cite{Golter2016}. Inspired by these discoveries,
we theoretically propose a novel method to polarize a $^{13}\mathrm{C}$
nuclear spin in the vicinity of an NV center by the dark state.
In order to provide an effective guidance for the experimental realization,
we performed an analytical analysis on the probability of the nuclear-spin
polarized state by the Schr\"{o}dinger equation with a non-Hermitian Hamiltonian.
It is shown that when the Rabi frequencies of the two pulses are equal,
the polarization is predict to reach the maximum at the given time.
Further numerical simulation demonstrates an anomalous effect that
the polarization of a nuclear spin with a smaller hyperfine interaction
can be even higher than a nuclear spin at the nearest neighbor site
due to the transverse hyperfine interaction.
Compared with the above methods, our scheme works effectively over a broad
range of magnetic field and only a few repetitions are required.

\begin{figure}
\includegraphics[bb=0bp 0bp 474bp 297bp,width=8.5cm]{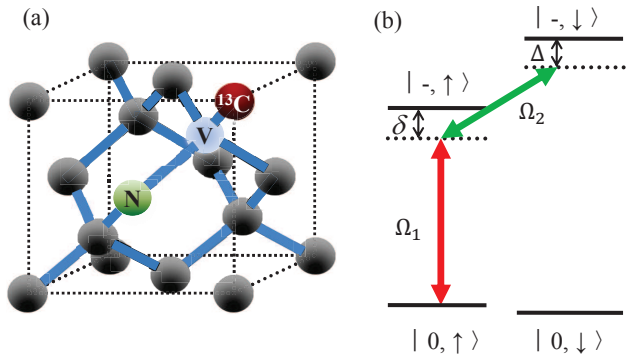}

\caption{Schematic of polarizing a $^{13}\mathrm{C}$ nuclear spin
in the vicinity of an NV center.
(a) The structure of an NV center. (b) Energy levels of the electron
and $^{13}\mathrm{C}$ nuclear spins under the hyperfine interaction.
Two pulses are simultaneously applied to induce the transitions of
$\vert0,\uparrow\rangle\leftrightarrow\vert-,\uparrow\rangle$ and
$\vert-,\uparrow\rangle\leftrightarrow\vert-,\downarrow\rangle$ with
Rabi frequencies $\Omega_{1}$ and $\Omega_{2}$, and
detunings $\delta$ and $\Delta$.\label{fig:nv-structure}}
\end{figure}

\section{Physical Setup}

\label{sec:PhysicalSetup}

As shown in Fig.~\ref{fig:nv-structure}(a), the NV center in diamond
with a $C_{3v}$ symmetry consists of a nitrogen atom associated with
a vacancy in an adjacent lattice site. For the negatively-charged
NV center with electron spin \emph{$S=1$}, the ground state is a
spin triplet state $^{3}\mathrm{A}$, with a zero-field splitting
$D=2.87$ GHz \cite{Grezes2014} between spin sublevels $m_{s}=0$
and $m_{s}=\pm1$ due to the spin-spin interaction. In this article,
we consider a first-shell $^{13}\mathrm{C}$ nuclear spin coupled
with the electronic spin of an NV center. As a result, there is a
strong hyperfine coupling $A_{\Vert}=130$ MHz \cite{Dreau2012} between
the nuclear and electronic spins. Figure~\ref{fig:nv-structure}(b)
shows the energy-level diagram of the ground-state hyperfine structure
associated with a nearby $^{13}\mathrm{C}$ nuclear spin. We label
the states of this bipartite system as $\vert m_{s},\uparrow\!\bigl\rangle$
and $\bigr\vert m_{s},\downarrow\!\bigl\rangle$, where $\bigr\vert\!\uparrow\!\bigl\rangle$
and $\bigr\vert\!\downarrow\!\bigl\rangle$ are the nuclear spin states.

We apply a weak static magnetic field $B_{z}<673$ G along the NV principle
axis by a permanent magnet. The total Hamiltonian of the electron-spin
ground state and a nearby $^{13}\mathrm{C}$ nuclear spin reads \cite{Shim2013}
\begin{eqnarray}
H_{F} & = & DS_{z}^{2}+\gamma_{e}B_{z}S_{z}+\gamma_{c}B_{z}I_{z}+A_{\Vert}S_{z}I_{z}\nonumber \\
 &  & +A_{\perp}(S_{x}I_{x}+S_{y}I_{y}).\label{eq:H}
\end{eqnarray}
Here, $S_{\alpha}$ and $I_{\alpha}$ ($\alpha=x,y,z$) are respectively
the electronic and nuclear spin operators. The first term stands for
the zero-field splitting of the electronic ground state. The following
two terms $\gamma_{e}B_{z}S_{z}$ and $\gamma_{c}B_{z}I_{z}$ are
the electronic and nuclear spin Zeeman energy splittings with the
electronic gyromagnetic ratio $\gamma_{e}=-1.76\times10^{11}$ rad
s$^{-1}$T$^{-1}$ \cite{Rugar2015} and the nuclear gyromagnetic
ratio $\gamma_{c}=6.73\times10^{7}$ rad s$^{-1}$T$^{-1}$ \cite{Zhao2011}.
The last two terms describe the hyperfine interaction between the
electron spin and the $^{13}\mathrm{C}$ nuclear spin, where $A_{\Vert}$
and $A_{\perp}$ are the longitudinal and transverse hyperfine interactions
respectively.

Due to the weak magnetic field strength, the difference between the
electronic Zeeman splitting and the zero-field splitting is much larger
than the transverse hyperfine interaction,
i.e. $\left\vert D-\gamma_{c}B_{z}\right\vert\gg A_{\perp}$.
Therefore the $S_{x}I_{x}$ and $S_{y}I_{y}$ terms of the hyperfine
interaction are sufficiently suppressed. In this case, the longitudinal
hyperfine interaction is taken into account and thus the secular approximation
is valid \cite{Childress2006,Neumann2008,Dreau2013,Ai2010}. In the
absence of time-varying magnetic fields, the Hamiltonian can be approximated
as
\begin{eqnarray}
H_{F}^{S} & \simeq & DS_{z}^{2}+\gamma_{e}B_{z}S_{z}+\gamma_{c}B_{z}I_{z}+A_{\Vert}S_{z}I_{z}.\label{eq:HF}
\end{eqnarray}

\section{Polarizing by Dark State}

\label{sec:Dynamics}

As shown in Fig.~\ref{fig:nv-structure}(b), the transition $\bigr\vert0,\uparrow\!\bigl\rangle$
$\leftrightarrow$ $\left\vert-,\uparrow\right\rangle $ is addressed
via a microwave pulse with Rabi frequency $\varOmega_{1}$ and driving
frequency $\omega_{A}=D-\gamma_{e}B_{z}-\delta-A_{\Vert}/2$, while
the transition $\bigr\vert-,\uparrow\!\bigl\rangle$ $\leftrightarrow$
$\left\vert-,\downarrow\right\rangle $ is driven via a radio-frequency
pulse with Rabi frequency $\varOmega_{2}$ and driving frequency $\omega_{B}=A_{\Vert}-\gamma_{c}B_{z}+\delta-\Delta$.
Thus, in the presence of the two pulses the whole Hamiltonian of the
system reads $H_{M}=H_{F}+H_{I}$, where the interaction Hamiltonian
is

\begin{eqnarray}
H_{I} & = & \text{\textgreek{W}}_{1}e^{i\omega_{A}t}\bigr\vert0,\uparrow\!\bigl\rangle\bigr\langle-,\uparrow\bigr\vert+\text{\textgreek{W}}_{2}e^{i\omega_{B}t}\bigr\vert-,\uparrow\!\bigl\rangle\bigr\langle-,\downarrow\bigr\vert+\mathrm{h.c.}\nonumber \\
\end{eqnarray}

Hereafter, we will demonstrate polarizing the nuclear spin by the
dark state. The nuclear and electronic spins are initially in a product
state \cite{Dutt2007,Neumann2010,Shim2013}. In each cycle there are
two steps. First of all, the system evolves under the influence of
the simultaneous microwave and radio-frequency pulses. In the second
step, the electronic and nuclear spins are decoupled by a 532-nm optical
pumping which re-initializes the electron spin in its ground state
$\bigr\vert0\bigl\rangle$, while the state of the nuclear spin is unchanged,
 i.e. $\rho(t)\rightarrow\rho_e(0)\otimes\textrm{Tr}_e\rho(t)$ \cite{Dutt2007,Neumann2010}. Then, the above cycle is repeated until a high polarization
of the nuclear spin is reached.

The optical excitation with a 532-nm laser pulse leads to a strong
spin polarization into the $\bigr\vert0\bigl\rangle$ sublevel of the
ground state \cite{Doi2014}, which derives from spin-selective nonradiative
intersystem crossing to a metastable state between the ground and
excited triple states. In this sense, it is reasonable to choose the
electronic initial state $\rho_{e}(0)=\bigr\vert0\bigl\rangle\bigr\langle0\bigr\vert$.
Due to the small nuclear Zeeman energy splitting with respect to the
thermal energy, the nuclear spin is in the maximum mixed state $\rho_{c}(0)\!=\!(\bigr\vert\!\uparrow\!\bigl\rangle\bigr\langle\!\uparrow\bigr\vert+\bigr\vert\!\downarrow\!\bigl\rangle\bigr\langle\!\downarrow\!\vert)/2$.
When the electronic spin is populated in $m_{s}=0$, the hyperfine
interaction vanishes. Thus, the initial state of the total system
is given by
\begin{eqnarray}
\rho(0) & = & \rho_{e}(0)\otimes\rho_{c}(0).
\end{eqnarray}

Then, the microwave pulse and the radio-frequency pulse drive the
transitions $\bigr\vert0,\uparrow\!\bigl\rangle$ $\leftrightarrow$ $\left\vert-,\uparrow\right\rangle $
and $\bigr\vert-,\uparrow\!\bigl\rangle$ $\leftrightarrow$ $\left\vert-,\downarrow\right\rangle $
respectively. The total system evolves under the Hamiltonian $H_{M}=H_{F}+H_{I}$
for a time interval $t$. Since the total Hamiltonian $H_{M}$ is
time-dependent, it is transformed to the rotating frame defined by
$\bigr\vert\varPsi(t)^{R}\bigl\rangle=U^{\dagger}(t)\bigr\vert\varPsi(t)\bigl\rangle$
with $U(t)=\exp[-i(H_{F}^{S}-\delta\bigr\vert-,\uparrow\!\bigl\rangle\bigr\langle-,\uparrow\bigr\vert)t]$.
Here, $\bigr\vert\varPsi(t)^{R}\bigl\rangle$ satisfies the Schr\"{o}dinger
equation in the rotating frame with the effective Hamiltonian

\begin{eqnarray}
H_{M}^{R} & \equiv & U^{\dagger}(H_{F}^{S}+H_{I})U+i\dot{U^{\dagger}}U\nonumber \\
 & = & \delta\vert-,\uparrow\rangle\langle-,\uparrow\vert+\Delta\vert-,\downarrow\rangle\langle-,\downarrow\vert\nonumber \\
 &  & +\Omega_{1}\vert0,\uparrow\rangle\langle-,\uparrow\vert+\Omega_{2}\vert-,\uparrow\rangle\langle-,\downarrow\vert+\mathrm{h.c.}
\end{eqnarray}

Generally speaking, the quantum dynamics of the electron and nuclear
spins is subject to the noise, which can be described by the master
equation
\begin{equation}
\partial_{t}\rho=-i[H_{M},\rho]+\mathcal{L}\rho,
\end{equation}
where $\mathcal{L}\rho=\kappa(\bigr\vert-\bigl\rangle\bigr\langle-\bigr\vert\rho\bigr\vert-\bigl\rangle\bigr\langle-\bigr\vert-\frac{1}{2}\{\bigr\vert-\bigl\rangle\bigr\langle-\bigr\vert,\rho\})$
describes the decoherence induced by the bath with $\kappa$ being
the decoherence rate, $\{\cdot,\rho\}$ is the anti-commutator.Because $T_{1}$ is at least larger than
$T_{2}=\kappa^{-1}$ by one order \cite{Takahashi2008,Maze2008,Jarmola2012},
without loss of generality, we only take the pure-dephasing process
into consideration in our simulation.

When the decoherence is sufficiently slow as compared to the coherent processes
described by $H_{M}$, the total quantum dynamics including the decoherence
can be simulated by the Schr\"{o}dinger equation with a non-Hermitian
Hamiltonian $H=H_{M}^{R}-\frac{i}{2}\kappa\bigr\vert-\bigl\rangle\bigr\langle-\bigr\vert$
\cite{Dong2012,Ai2014}.
Because we apply two selective-resonance drivings to the NV center,
the state $\vert0,\downarrow\!\bigl\rangle$ is effectively decoupled
from the other three states. Therefore, hereafter we can separately
consider the quantum dynamics of an initial state $\vert\psi(0)\rangle=\vert0,\uparrow\rangle$
driven by two microwave pulses. As presented in Appendix~\ref{sec:appDS},
at any time the state of the system reads
\begin{align}
\vert\psi(t)\rangle & =\sum_{j=1}^{3}\frac{N_{j}e^{-ix_{j}t}}{\prod_{k\neq j}(x_{j}-x_{k})}\left\vert E_{j}\right\rangle ,
\end{align}
where the three eigen states of $H$ are
\begin{eqnarray}
\left\vert E_{j}\right\rangle  & \!=\! & \frac{1}{N_{j}}\{\text{\textgreek{W}}_{1}(x_{j}-\omega_{2})\left\vert-,\uparrow\right\rangle +\text{\textgreek{W}}_{1}\text{\textgreek{W}}_{2}\left\vert-,\downarrow\right\rangle \nonumber \\
 &  & +[(x_{j}-\omega_{1})(x_{j}-\omega_{2})-\text{\textgreek{W}}_{2}^{2}]\left\vert0,\uparrow\right\rangle \}
\end{eqnarray}
with $N_{j}$'s being the normalization constants, $x_{j}$'s being
the eigen energies, $\omega_{1}=\delta-i\kappa/2$, $\omega_{2}=\Delta-i\kappa/2$.
When we employ two strong drivings to polarize the nuclear spin, i.e.
$\omega_{1},\omega_{2}\ll\Omega_{1},\Omega_{2}$,
\begin{equation}
\left\vert E_{1}\right\rangle \simeq\frac{\text{\textgreek{W}}_{2}}{N_{1}}(-\text{\textgreek{W}}_{2}\left\vert0,\uparrow\right\rangle +\text{\textgreek{W}}_{1}\left\vert-,\downarrow\right\rangle )
\end{equation}
is the dark state because it is lack of the component of the lossy
intermediate state $\left\vert-,\uparrow\right\rangle $, while
\begin{align}
\left\vert E_{2}\right\rangle  & \simeq\frac{\text{\textgreek{W}}_{1}}{N_{2}}(\text{\textgreek{W}}_{1}\left\vert0,\uparrow\right\rangle +\Omega\left\vert-,\uparrow\right\rangle +\text{\textgreek{W}}_{2}\left\vert-,\downarrow\right\rangle ),\\
\left\vert E_{3}\right\rangle  & \simeq\frac{\text{\textgreek{W}}_{1}}{N_{3}}(\text{\textgreek{W}}_{1}\left\vert0,\uparrow\right\rangle -\Omega\left\vert-,\uparrow\right\rangle +\text{\textgreek{W}}_{2}\left\vert-,\downarrow\right\rangle )
\end{align}
are the bright states suffering from relaxation. In this case, the
state of system is simplified as
\begin{eqnarray}
\vert\psi(t)\rangle & = & \frac{\text{\textgreek{W}}_{1}}{\sqrt{2}\Omega}e^{-i\frac{1}{2}(\omega_{1}+\frac{\text{\textgreek{W}}_{2}^{2}}{\text{\textgreek{W}}^{2}}\omega_{2})t}(e^{-i\text{\textgreek{W}}t}\left\vert E_{2}\right\rangle +e^{i\text{\textgreek{W}}t}\left\vert E_{3}\right\rangle )\nonumber \\
 &  & -\frac{\text{\textgreek{W}}_{2}}{\Omega}e^{-i\frac{\text{\textgreek{W}}_{1}^{2}}{\text{\textgreek{W}}^{2}}\omega_{2}t}\left\vert E_{1}\right\rangle
\end{eqnarray}
with
\begin{align}
\Omega & =\sqrt{\Omega_{1}^{2}+\Omega_{2}^{2}}.
\end{align}

\begin{figure}
\includegraphics[clip,width=9cm]{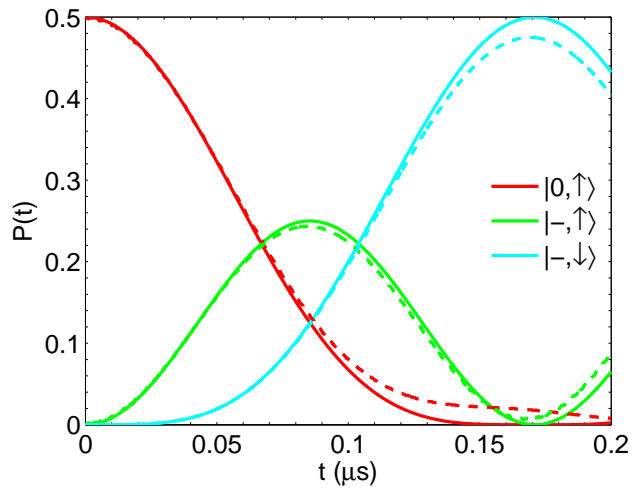}

\caption{The population dynamics of $\vert0,\uparrow\rangle$ (red), $\vert-,\uparrow\rangle$ (green), and $\vert-,\downarrow\rangle$ (blue) by the Schr\"{o}dinger equation (solid
lines) with a non-Hermitian Hamiltonian and the exact master equation (dashed
lines). The population dynamics are drawn with $A_{\Vert}=130$ MHz
\cite{Smeltzer2011}, $\delta=\Delta=0$, $\Omega_{1}=\Omega_{2}=13$ MHz,
$\kappa=1/58$ MHz \cite{Kennedy2003}.\label{fig:nonHermVsMasEq}}
\end{figure}

By solving the Schr\"{o}dinger equation with a non-Hermitian Hamiltonian,
we numerically simulate the population dynamics of all three states
for the resonance case, i.e. $\delta=\Delta=0$, as shown by the solid
lines in Fig.~\ref{fig:nonHermVsMasEq}. In one cycle, almost 100\%
population in $\bigr\vert0,\uparrow\!\bigl\rangle$ can be coherently
transferred to $\bigr\vert-,\downarrow\!\bigl\rangle$ even in the presence
of noise. In order to derive the non-Hermitian Hamiltonian, several
approximations have been utilized, i.e. dropping the quantum jump
terms in the master equation, and ignoring the transverse hyperfine
interactions, and disregarding transitions due to large-detuning condition.
In order to validate these approximations, we also present the numerical
simulation with the dashed lines in Fig.~\ref{fig:nonHermVsMasEq}
by the exact master equation without the above approximations. Obviously,
the differences between the two approaches are relative small and
thus it is valid to simulate the quantum dynamics of the nuclear-spin
polarization in the presence of noise.

\begin{figure}
\includegraphics[clip,width=9cm]{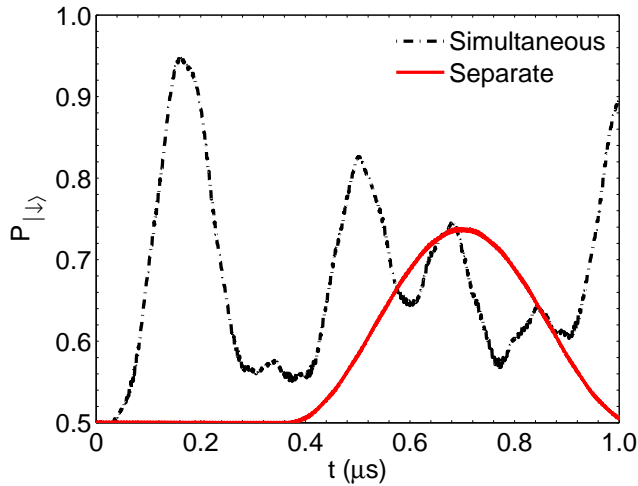}

\caption{Comparison between the probabilities of the nuclear spin in the $\vert\downarrow\rangle$
vs time by applying two simultaneous pulses
(black dash-dotted line) and two separate pulses (red solid line).
In the simultaneous case, $\Omega_{1}=\Omega_{2}=13$ MHz, while $\Omega_{1}=\Omega_{2}=4.3$ MHz
\cite{Shim2013} in the separate case. Other parameters are the same
in both cases, i.e. $A_{\Vert}=130$ MHz \cite{Smeltzer2011}, $\delta=\Delta=0$,
$\kappa=1/58$ MHz \cite{Kennedy2003}.\label{fig:SimulVsSep}}
\end{figure}

By means of the Schr\"{o}dinger equation with a non-Hermitian Hamiltonian,
we can effectively analyze the effects of parameters on the nuclear-spin
polarization and obtain a set of optimal parameters to guide the experiment
for different conditions of nuclear spins. In the previous experimental realizations, cf. Refs.~\cite{Childress2006,Dutt2007,Neumann2008,Taminiau2014,Shim2013},
two $\pi$-pulses are sequentially applied to swap the electron-spin
polarization into the nuclear-spin polarization. In our proposal,
because we simultaneously apply two balanced pulses to make use of
the dark state to avoid the noise suffered by the intermediate state,
the nuclear-spin polarization in our proposal is much
higher than theirs. As shown in Fig.~\ref{fig:SimulVsSep}, the polarization
for our proposal is further optimized by choosing optimal parameters
to reduce the swapping time, i.e. $0.90$ vs $0.48$.

\begin{figure*}
\includegraphics[width=6cm]{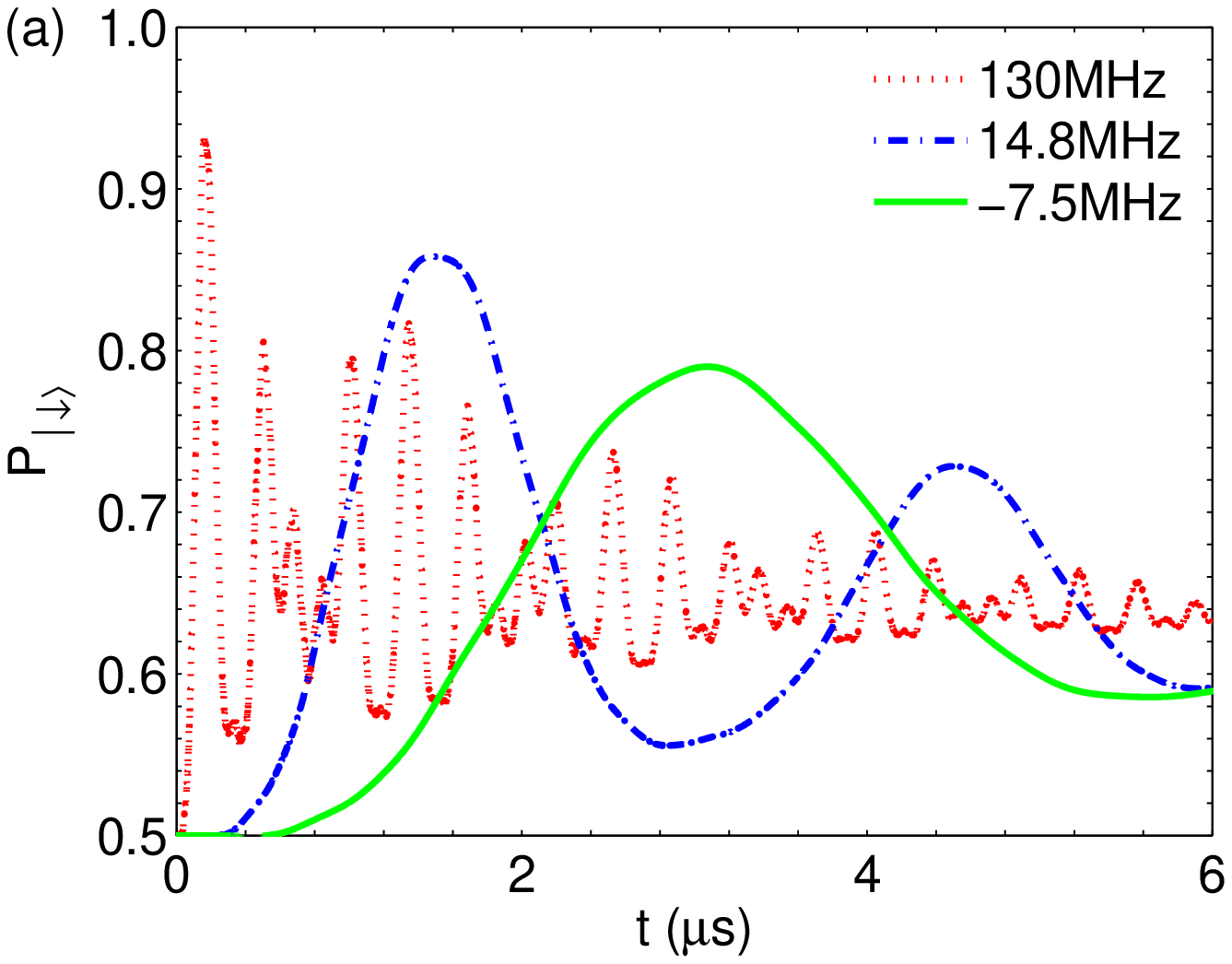}\includegraphics[width=6cm]{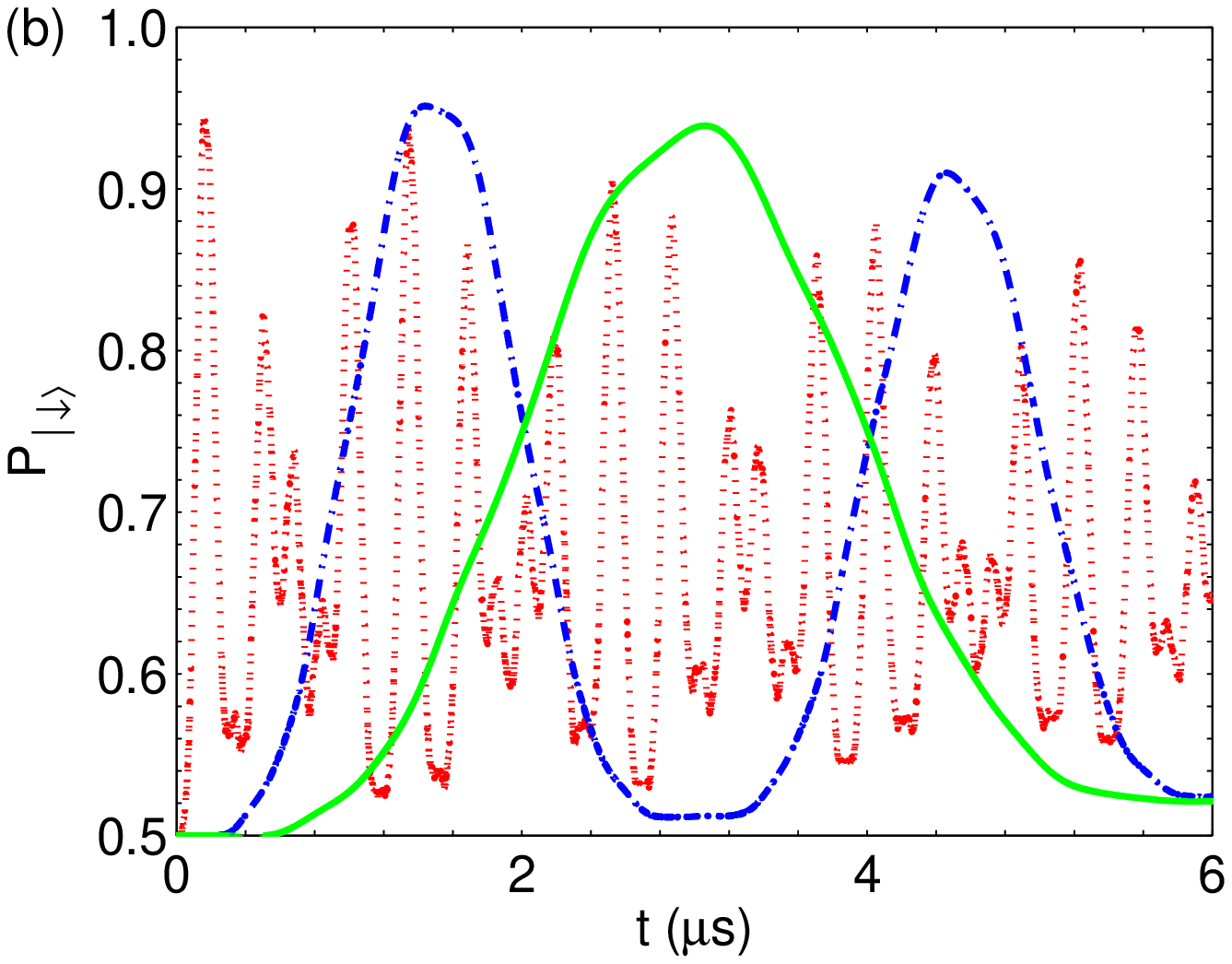}\includegraphics[width=6cm]{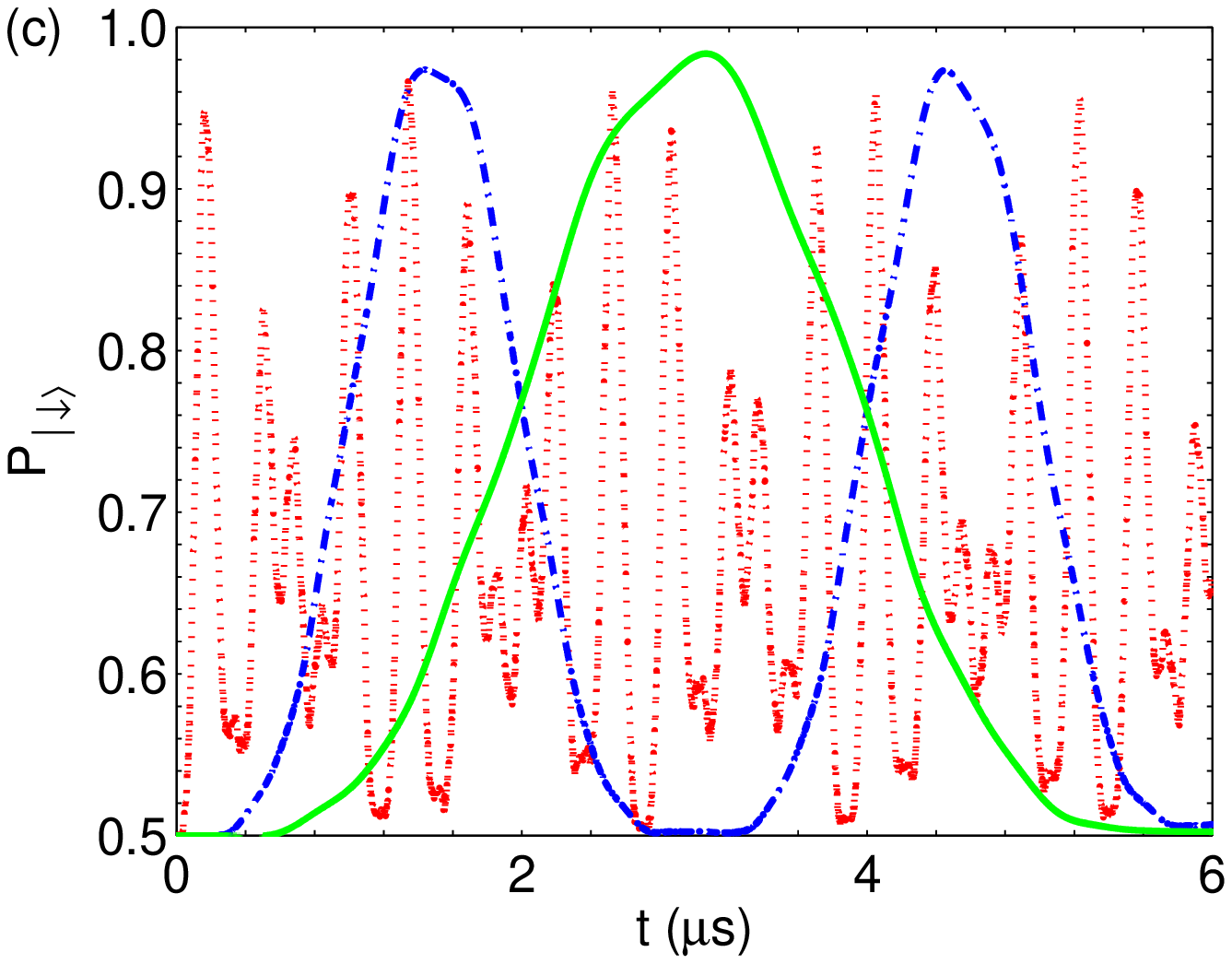}\caption{The fidelity of nuclear spin in $\bigr\vert\downarrow\bigl\rangle$ vs
$A_{\Vert}$ = $A_{\perp}$ for different noise strength: (a) $\kappa=1$ MHz,
(b) $\kappa=1/5.8$ MHz, (c) $\kappa=1/58$ MHz \cite{Kennedy2003}.
Red dotted line is $A_{\Vert}=130$ MHz \cite{Smeltzer2011}, and blue
dash-dotted line is $A_{\Vert}=14.8$ MHz \cite{Smeltzer2011}, and
green solid line is $A_{\Vert}=-7.5$ MHz \cite{Smeltzer2011}. Other
parameters are the same as in Fig.~\ref{fig:nonHermVsMasEq}.}
\label{fig:PopVs_A_kappa}
\end{figure*}

The swapping of electron-spin polarization into nuclear-spin polarization
is intrinsically a quantum state transfer process from $\vert0,\uparrow\rangle$
to $\vert-,\downarrow\rangle$ via a lossy state $\vert-,\uparrow\rangle$.
Intuitively, the fidelity of state transfer is subject to the noise
strength. In Fig.~\ref{fig:nonHermVsMasEq}(a), we explore the noise's
effect on the fidelity of nuclear spin in the state $\vert\!\downarrow\rangle$
with $\kappa=1$MHz for different nuclear spins, i.e. different hyperfine interactions.
For a nuclear spin in the first shell, the probability in $\vert\!\downarrow\rangle$
vs time manifests a damped vibration due to couplings to the environment.
The maximum fidelity is more than 0.9 around the first peak at $t=\pi/\sqrt{2}\Omega_{1}\simeq0.17\mu$s
as a longer pulse duration yields more loss. When a nuclear spin in
the second shell is to be polarized, the maximum fidelity is about
0.85, less than that for the nuclear spin in the first shell as the
Rabi frequency is smaller due to a weaker hyperfine interaction. If
we choose a nuclear spin even further apart from the NV center, e.g.
$A_{\Vert}=-7.5$ MHz, the maximum fidelity observably declines to 0.65.
Because Rabi frequencies are limited by the large-detuning condition,
the descending of maximum fidelity along with reducing of hyperfine
interaction results from the increasing pulse duration. When the pure-dephasing
rate $\kappa$ is reduced from $1$ MHz to $1/5.8$ MHz, e.g. Fig.~\ref{fig:PopVs_A_kappa}(a)
vs Fig.~\ref{fig:PopVs_A_kappa}(b), an anomalous phenomena occurs,
the maximum fidelity achieved for the nuclear spin in the second shell
is a little bit larger than that for the nuclear spin in the first
shell. That is because the transverse hyperfine interaction provides
an additional pathway for the intermediate state $\vert-,\uparrow\rangle$
to the nuclear-spin polarized state $\vert0,\downarrow\rangle$. If
the noise strength is further suppressed to $\kappa=1/58$ MHz, the
first peak for a nuclear spin with $A_{\Vert}=-7.5$ MHz rises although
more time is required for the evolution.

\section{Discussion and Conclusion}

The initialization of the nuclear spin is critical to the subsequent
quantum information storage and processing. Facilitated by the strong
hyperfine interaction with the electron spin, the nuclear spin
in the vicinity of an NV center can be polarized by
the swapping of the electron-spin polarization. And the swapping process
is intrinsically a quantum-state process via a lossy intermediate state.
In this paper, we propose to polarize the $^{13}\mathrm{C}$ nuclear
spin coupled to the electron spin of an NV center through the dark
state. Our simulation demonstrates that the nuclear-spin polarization
can reach more than 96.7\% for the next-next-nearest neighbor site.
In the following, we will discuss the feasibility in the experiment
and advantages of this proposal.

In theory, in the case of stronger Rabi frequencies $\Omega_{1}$
and $\Omega_{2}$, it takes a shorter time $t$ for the nuclear spin
to reach the maximum polarization. In our scheme, the magnitudes of
$\Omega_{1}$ and $\Omega_{2}$ are limited by the selective-excitation
condition. The driving frequency $\omega_{A}$ is set to be in close
resonance with the transition $\bigr\vert0,\uparrow\bigl\rangle\rightleftharpoons\bigr\vert-,\uparrow\bigl\rangle$
, i.e. $\Omega_{1}\gtrsim\delta$. On the other hand, the level spacing
between $\bigr\vert0,\downarrow\bigl\rangle$ and $\bigr\vert-,\downarrow\bigl\rangle$
is $D-\gamma_{e}B_{z}+A_{\Vert}/2$. To selectively address the transition
between $\bigr\vert0,\uparrow\bigl\rangle$ and $\bigr\vert-,\uparrow\bigl\rangle$,
the Rabi frequency $\Omega_{1}$ must satisfy the large-detuning condition,
i.e. $\Omega_{1}\ll\delta+A_{\Vert}$. In the same way, we can deduce
that $\delta-\Delta\lesssim\Omega_{2}\ll A_{\Vert}-\delta+\Delta$.
To be specific, because the hyperfine interaction between the electronic
spin and the $^{13}\mathrm{C}$ nuclear spin in the first shell $A_{\Vert}$
is $130$MHz, the Rabi frequencies can be no more than $13$MHz. Furthermore,
in order to validate the secular approximation, the magnetic field
strength and the hyperfine interaction should fulfill the requirement
$D-\gamma_{e}B_{z}+\gamma_{c}B_{z}-\frac{1}{2}A_{\Vert}\gg A_{\perp}/\sqrt{2}$.
In other words, $B_{z}\leq673$G for the case with a $^{13}\mathrm{C}$
nuclear spin in the first shell.

Compared with Refs.~\cite{Jacques2009,Wang2013}, our scheme does
not require a specific magnetic field to result in a level anti-crossing
in the ground or excited states. Besides, our proposal is not sensitive
to the inclination of applied magnetic field. In Refs.~\cite{Shim2013,Dutt2007},
the initialization of nuclear spin with fidelity 85\% is realized
by mapping the electronic spin polarization into the nuclear spin
state with a weak hyperfine interaction. Furthermore, our proposal
can reach 96\% in a weaker hyperfine interaction. Since about 10 repetitions
are required in our proposal, it is more convenient than the single-shot
readout approach in Refs.~\cite{Neumann2010,Jiang2009}.

\begin{acknowledgments}
We thank stimulating discussions with Y. D. Wang and H. Dong. F.-G.D. was supported
by the National Natural Science Foundation of China under Grant No.~11474026
and the Fundamental Research Funds for the Central Universities under
Grant No.~2015KJJCA01. M.Z. was supported by the National Natural Science
Foundation of China under Grant No.~11475021. J.-M.C. is supported
by the National Natural Science Foundation of China (Grant No.11574103),
and the National Young 1000 Talents Plan. Q.A. was supported by the
National Natural Science Foundation of China under Grant No.~11505007,
and the Open Research Fund Program of the State Key Laboratory of
Low-Dimensional Quantum Physics, Tsinghua University Grant No.~KF201502.
\end{acknowledgments}

\appendix
\section{Quantum Dynamics and Dark State}

\label{sec:appDS}

In the basis of \{$\vert0,\uparrow\rangle,\vert-,\uparrow\rangle,\vert-,\downarrow\rangle$\},
the Hamiltonian is written in the matrix form as
\begin{equation}
H=\begin{pmatrix}0 & \Omega_{1} & 0\\
\Omega_{1} & \omega_{1} & \Omega_{2}\\
0 & \Omega_{2} & \omega_{2}
\end{pmatrix},\label{eq:nonHermitianH}
\end{equation}
where
\begin{align}
\omega_{1} & =\delta-i\kappa/2,\\
\omega_{2} & =\Delta-i\kappa/2.
\end{align}

We consider the time evolution of an initial state $\vert\psi(0)\rangle=\vert0,\uparrow\rangle$.
At any time, the state reads
\begin{equation}
\vert\psi(t)\rangle=u(t)\vert0,\uparrow\rangle+v(t)\vert-,\uparrow\rangle+w(t)\vert-,\downarrow\rangle.
\end{equation}
According to Schr\"{o}dinger equation, we obtain a set of differential
equations for the probability amplitudes as\begin{subequations}
\begin{align}
i\dot{u}(t) & =\text{\textgreek{W}}_{1}v(t),\\
i\dot{v}(t) & =\text{\textgreek{W}}_{1}u(t)+\omega_{1}v(t)+\text{\textgreek{W}}_{2}w(t),\\
\dot{iw}(t) & =\text{\textgreek{W}}_{2}v(t)+\omega_{2}w(t),
\end{align}
\end{subequations}with the initial condition $u(0)=1$, $v(0)=w(0)=0$.
By Laplace transformation, $\widetilde{\alpha}(p)=\int_{0}^{\infty}\alpha(t)e^{-pt}dt$
($\alpha=u,v,w$),\begin{subequations}
\begin{align}
i[p\widetilde{u}(p)-1] & =\text{\textgreek{W}}_{1}\widetilde{v}(p),\\
ip\widetilde{v}(p) & =\text{\textgreek{W}}_{1}\widetilde{u}(p)+\omega_{1}\widetilde{v}(p)+\text{\textgreek{W}}_{2}\widetilde{w}(p),\\
ip\widetilde{w}(p) & =\text{\textgreek{W}}_{2}\widetilde{v}(p)+\omega_{2}\widetilde{w}(p).
\end{align}
\end{subequations}or equivalently in the matrix form
\begin{equation}
\left[\begin{array}{ccc}
-ip & \Omega_{1} & 0\\
\text{\textgreek{W}}_{1} & \omega_{1}-ip & \text{\textgreek{W}}_{2}\\
0 & \text{\textgreek{W}}_{2} & \omega_{2}-ip
\end{array}\right]\left[\begin{array}{c}
\widetilde{u}(p)\\
\widetilde{v}(p)\\
\widetilde{w}(p)
\end{array}\right]=\left[\begin{array}{c}
-i\\
0\\
0
\end{array}\right].
\end{equation}

We define\begin{subequations}

\begin{align}
\textrm{det}D & \equiv\left\vert\begin{array}{ccc}
-ip & \text{\textgreek{W}}_{1} & 0\\
\text{\textgreek{W}}_{1} & \omega_{1}-ip & \text{\textgreek{W}}_{2}\\
0 & \text{\textgreek{W}}_{2} & \omega_{2}-ip
\end{array}\right\vert\nonumber \\
 & =(x_{1}-ip)(x_{2}-ip)(x_{3}-ip),
\end{align}

\begin{align}
\textrm{det}D_{1} & \equiv\left\vert\begin{array}{ccc}
-i & \text{\textgreek{W}}_{1} & 0\\
0 & \omega_{1}-ip & \text{\textgreek{W}}_{2}\\
0 & \text{\textgreek{W}}_{2} & \omega_{2}-ip
\end{array}\right\vert\nonumber \\
 & =-i(\omega_{1}-ip)(\omega_{2}-ip)-(-i)\text{\textgreek{W}}_{2}^{2},
\end{align}

\begin{equation}
\textrm{det}D_{2}\equiv\left\vert\begin{array}{ccc}
-ip & -i & 0\\
\text{\textgreek{W}}_{1} & 0 & \text{\textgreek{W}}_{2}\\
0 & 0 & \omega_{2}-ip
\end{array}\right\vert=i\text{\textgreek{W}}_{1}(\omega_{2}-ip),
\end{equation}

\begin{equation}
\textrm{det}D_{3}\equiv\left\vert\begin{array}{ccc}
-ip & \text{\textgreek{W}}_{1} & -i\\
\text{\textgreek{W}}_{1} & \omega_{1}-ip & 0\\
0 & \text{\textgreek{W}}_{2} & 0
\end{array}\right\vert=-i\text{\textgreek{W}}_{1}\text{\textgreek{W}}_{2},
\end{equation}
\end{subequations}where $x_{j}$'s are the eigen energies of Hamiltonian~(\ref{eq:nonHermitianH}),
which will be determined later. And thus we have\begin{subequations}
\begin{align}
\widetilde{u}(p) & =\frac{\textrm{det}D_{1}}{\textrm{det}D}=-i\frac{(\omega_{1}-ip)(\omega_{2}-ip)-\text{\textgreek{W}}_{2}^{2}}{(x_{1}-ip)(x_{2}-ip)(x_{3}-ip)},\\
\widetilde{v}(p) & =\frac{\textrm{det}D_{2}}{\textrm{det}D}=-i\frac{\text{\textgreek{W}}_{1}(ip-\omega_{2})}{(x_{1}-ip)(x_{2}-ip)(x_{3}-ip)},\\
\widetilde{w}(p) & =\frac{\textrm{det}D_{3}}{\textrm{det}D}=-i\frac{\text{\textgreek{W}}_{1}\text{\textgreek{W}}_{2}}{(x_{1}-ip)(x_{2}-ip)(x_{3}-ip)}.
\end{align}
\end{subequations}Furthermore, by inverse Laplace transformation,
$\alpha(t)=(2\pi i)^{-1}\int_{\sigma+i\infty}^{\sigma+i\infty}\widetilde{\alpha}(p)e^{pt}dp$
($\alpha=u,v,w$), we obtain the probability amplitudes as\begin{subequations}
\begin{align}
u(t) & =\sum_{j=1}^{3}\frac{(x_{j}-\omega_{1})(x_{j}-\omega_{2})-\text{\textgreek{W}}_{2}^{2}}{\prod_{k\neq j}(x_{j}-x_{k})}e^{-ix_{j}t},\\
v(t) & =\sum_{j=1}^{3}\frac{\text{\textgreek{W}}_{1}(x_{j}-\omega_{2})}{\prod_{k\neq j}(x_{j}-x_{k})}e^{-ix_{j}t},\\
w(t) & =\sum_{j=1}^{3}\frac{\text{\textgreek{W}}_{1}\text{\textgreek{W}}_{2}}{\prod_{k\neq j}(x_{j}-x_{k})}e^{-ix_{j}t}.
\end{align}
\end{subequations}

According to Schr\"{o}dinger equation, the eigen energies $x_{j}$'s
are the solutions to the following equation

\begin{equation}
x[x^{2}-(\omega_{1}+\omega_{2})x+\omega_{1}\omega_{2}-\text{\textgreek{W}}^{2}]+\text{\textgreek{W}}_{1}^{2}\omega_{2}=0,\label{eq:ExEigenEq}
\end{equation}
where
\begin{equation}
\text{\textgreek{W}}^{2}=\text{\textgreek{W}}_{1}^{2}+\text{\textgreek{W}}_{2}^{2}.
\end{equation}

When $\omega_{2}$ is small, assuming
\begin{align}
x_{j}\text{\ensuremath{\simeq}} & x_{0j}+A_{j}\omega_{2},
\end{align}
we can obtain the approximate solutions by the perturbation theory.
The zeroth-order terms are determined by

\begin{equation}
x_{0j}[x_{0j}^{2}-(\omega_{1}+\omega_{2})x_{0j}+\omega_{1}\omega_{2}-\text{\textgreek{W}}^{2}]=0,
\end{equation}
where\begin{subequations}

\begin{align}
x_{01} & =0,\\
x_{02} & =\omega_{+},\\
x_{03} & =\omega_{-},\\
\omega_{\pm} & =\frac{1}{2}[(\omega_{1}+\omega_{2})\pm\sqrt{(\omega_{1}-\omega_{2})^{2}+4\text{\textgreek{W}}^{2}}].
\end{align}
\end{subequations}

Therefore, Eq.~(\ref{eq:ExEigenEq}) can be rewritten as
\begin{equation}
x(x-\omega_{+})(x-\omega_{-})+\text{\textgreek{W}}_{1}^{2}\omega_{2}=0.\label{eq:AppEigenEq}
\end{equation}
By inserting $x_{1}=A_{1}\omega_{2}$ into Eq.~(\ref{eq:AppEigenEq}),
to the zeroth order of $\omega_{2}$, we obtain

\begin{equation}
A_{1}=-\frac{\text{\textgreek{W}}_{1}^{2}}{\omega_{+}\omega_{-}}\simeq\frac{\text{\textgreek{W}}_{1}^{2}}{\text{\textgreek{W}}^{2}}.
\end{equation}
By inserting $x_{2}=\omega_{+}+A_{2}\omega_{2}$ into Eq.~(\ref{eq:AppEigenEq}),
to the zeroth order of $\omega_{2}$, we obtain

\begin{equation}
A_{2}=-\frac{\text{\textgreek{W}}_{1}^{2}}{\omega_{+}(\omega_{+}-\omega_{-})}\simeq-\frac{\text{\textgreek{W}}_{1}^{2}}{2\text{\textgreek{W}}^{2}}.
\end{equation}
By inserting $x_{3}=\omega_{-}+A_{3}\omega_{2}$ into Eq.~(\ref{eq:AppEigenEq}),
to the zeroth order of $\omega_{2}$, we obtain

\begin{equation}
A_{3}=\frac{\text{\textgreek{W}}_{1}^{2}}{\omega_{-}(\omega_{+}-\omega_{-})}\simeq-\frac{\text{\textgreek{W}}_{1}^{2}}{2\text{\textgreek{W}}^{2}}.
\end{equation}
The eigen states are
\begin{eqnarray}
\left\vert E_{i}\right\rangle  & \!=\! & \frac{1}{N_{i}}\{[(x_{i}-\omega_{1})(x_{i}-\omega_{2})-\text{\textgreek{W}}_{2}^{2}]\left\vert0,\uparrow\right\rangle \nonumber \\
 &  & +\text{\textgreek{W}}_{1}(x_{i}-\omega_{2})\left\vert-,\uparrow\right\rangle +\text{\textgreek{W}}_{1}\text{\textgreek{W}}_{2}\left\vert-,\downarrow\right\rangle \},
\end{eqnarray}
where the normalization constants are
\begin{eqnarray}
N_{i}^{2} & = & \left\vert(x_{i}-\omega_{1})(x_{i}-\omega_{2})-\text{\textgreek{W}}_{2}^{2}\right\vert^{2}+\left\vert\text{\textgreek{W}}_{1}(x_{i}-\omega_{2})\right\vert^{2}\nonumber \\
 &  & +\text{\ensuremath{\left\vert\text{\textgreek{W}}_{1}\text{\textgreek{W}}_{2}\right\vert^{2}}.}
\end{eqnarray}

In the eigen basis, the time evolution of the initial state $\vert\psi(0)\rangle=\left\vert0,\uparrow\right\rangle $
is
\begin{align}
\vert\psi(t)\rangle & =\sum_{j=1}^{3}\frac{N_{j}e^{-ix_{j}t}}{\prod_{k\neq j}(x_{j}-x_{k})}\left\vert E_{j}\right\rangle .
\end{align}
When $\omega_{1},\omega_{2}\ll\Omega_{1},\Omega_{2}$, to the first
order of $\omega_{j}$'s, we have
\begin{align}
\text{ \textgreek{w}}_{\pm} & =\frac{1}{2}\{(\omega_{1}+\omega_{2})\pm\sqrt{4\text{\textgreek{W}}^{2}[1+\frac{(\omega_{1}-\omega_{2})^{2}}{4\text{\textgreek{W}}^{2}}]}\}\nonumber \\
 & \simeq\frac{1}{2}\{(\omega_{1}+\omega_{2})\pm2\text{\textgreek{W}}[1+\frac{(\omega_{1}-\omega_{2})^{2}}{4\text{\textgreek{W}}^{2}}]\}\nonumber \\
 & \simeq\frac{1}{2}[(\omega_{1}+\omega_{2})\pm2\text{\textgreek{W}}].
\end{align}
The eigen energies are approximated to the first order of $\omega_{j}$'s
as\begin{subequations}
\begin{align}
x_{1} & \simeq\frac{\text{\textgreek{W}}_{1}^{2}}{\frac{1}{4}[4\text{\textgreek{W}}^{2}-(\omega_{1}+\omega_{2})^{2}]}\omega_{2}\nonumber \\
 & \simeq\frac{\text{\textgreek{W}}_{1}^{2}}{\text{\textgreek{W}}^{2}}\omega_{2},
\end{align}
\begin{align}
x_{2} & \simeq\frac{1}{2}[(\omega_{1}+\omega_{2})+2\text{\textgreek{W}}]-\frac{\text{\textgreek{W}}_{1}^{2}}{[(\omega_{1}+\omega_{2})+2\text{\textgreek{W}}]\text{\textgreek{W}}}\omega_{2}\nonumber \\
 & \simeq\text{\textgreek{W}}+\frac{1}{2}(\omega_{1}+\frac{\text{\textgreek{W}}_{2}^{2}}{\text{\textgreek{W}}^{2}}\omega_{2}),
\end{align}
\begin{align}
x_{3} & \simeq\frac{1}{2}[(\omega_{1}+\omega_{2})-2\text{\textgreek{W}}]+\frac{\text{\textgreek{W}}_{1}^{2}}{[(\omega_{1}+\omega_{2})-2\text{\textgreek{W}}]\text{\textgreek{W}}}\omega_{2}\nonumber \\
 & \simeq-\text{\textgreek{W}}+\frac{1}{2}(\omega_{1}+\frac{\text{\textgreek{W}}_{2}^{2}}{\text{\textgreek{W}}^{2}}\omega_{2}).
\end{align}
\end{subequations}Furthermore, the probability amplitudes can be
obtained with the coefficients to the zeroth order of $\omega_{j}$'s
and the phases to the first order of $\omega_{j}$'s as\begin{widetext}
\begin{align}
u(t) & =\frac{(x_{1}-\omega_{1})(x_{1}-\omega_{2})-\text{\textgreek{W}}_{2}^{2}}{(x_{1}-x_{2})(x_{1}-x_{3})}e^{-ix_{1}t}+\frac{(x_{2}-\omega_{1})(x_{2}-\omega_{2})-\text{\textgreek{W}}_{2}^{2}}{(x_{2}-x_{1})(x_{2}-x_{3})}e^{-ix_{2}t}+\frac{(x_{3}-\omega_{1})(x_{3}-\omega_{2})-\text{\textgreek{W}}_{2}^{2}}{(x_{3}-x_{1})(x_{3}-x_{2})}e^{-ix_{3}t}\nonumber \\
 & =\frac{(0-\omega_{1})(0-\omega_{2})-\text{\textgreek{W}}_{2}^{2}}{(0-\omega_{+})(0-\omega_{-})}e^{-i\frac{\text{\textgreek{W}}_{1}^{2}}{\text{\textgreek{W}}^{2}}\omega_{2}t}+\frac{(\omega_{+}-\omega_{1})(\omega_{+}-\omega_{2})-\text{\textgreek{W}}_{2}^{2}}{(\omega_{+}-0)(\omega_{+}-\omega_{-})}e^{-i[\text{\textgreek{W}}+\frac{1}{2}(\omega_{1}+\frac{\text{\textgreek{W}}_{2}^{2}}{\text{\textgreek{W}}^{2}}\omega_{2})]t}\nonumber \\
 & +\frac{(\omega_{-}-\omega_{1})(\omega_{-}-\omega_{2})-\text{\textgreek{W}}_{2}^{2}}{(\omega_{-}-0)(\omega_{-}-\omega_{+})}e^{-i[-\text{\textgreek{W}}+\frac{1}{2}(\omega_{1}+\frac{\text{\textgreek{W}}_{2}^{2}}{\text{\textgreek{W}}^{2}}\omega_{2})]t}\nonumber \\
 & =\frac{\text{\textgreek{W}}_{2}^{2}}{\text{\textgreek{W}}^{2}}e^{-i\frac{\text{\textgreek{W}}_{1}^{2}}{\text{\textgreek{W}}^{2}}\omega_{2}t}+\frac{\text{\textgreek{W}}^{2}-\text{\textgreek{W}}_{2}^{2}}{2\text{\textgreek{W}}^{2}}e^{-i[\text{\textgreek{W}}+\frac{1}{2}(\omega_{1}+\frac{\text{\textgreek{W}}_{2}^{2}}{\text{\textgreek{W}}^{2}}\omega_{2})]t}+\frac{\text{\textgreek{W}}^{2}-\text{\textgreek{W}}_{2}^{2}}{-2\text{\textgreek{W}}^{2}}e^{-i[-\text{\textgreek{W}}+\frac{1}{2}(\omega_{1}+\frac{\text{\textgreek{W}}_{2}^{2}}{\text{\textgreek{W}}^{2}}\omega_{2})]t}\nonumber \\
 & =\frac{\text{\textgreek{W}}_{2}^{2}}{\text{\textgreek{W}}^{2}}e^{-i\frac{\text{\textgreek{W}}_{1}^{2}}{\text{\textgreek{W}}^{2}}\omega_{2}t}-i\frac{\text{\textgreek{W}}_{1}^{2}}{\text{\textgreek{W}}^{2}}e^{-i\frac{1}{2}(\omega_{1}+\frac{\text{\textgreek{W}}_{2}^{2}}{\text{\textgreek{W}}^{2}}\omega_{2})t}\sin\text{\textgreek{W}}t,
\end{align}
\begin{align}
v(t) & =\frac{\text{\textgreek{W}}_{1}(x_{1}-\omega_{2})}{(x_{1}-x_{2})(x_{1}-x_{3})}e^{-ix_{1}t}+\frac{\text{\textgreek{W}}_{1}(x_{2}-\omega_{2})}{(x_{2}-x_{1})(x_{2}-x_{3})}e^{-ix_{2}t}+\frac{\text{\textgreek{W}}_{1}(x_{3}-\omega_{2})}{(x_{3}-x_{1})(x_{3}-x_{2})}e^{-ix_{3}t}\nonumber \\
 & =\frac{\text{\textgreek{W}}_{1}(0-\omega_{2})}{(0-\text{\textgreek{W}})(0+\text{\textgreek{W}})}e^{-i\frac{\text{\textgreek{W}}_{1}^{2}}{\text{\textgreek{W}}^{2}}\omega_{2}t}+\frac{\text{\textgreek{W}}_{1}(\text{\textgreek{W}}-\omega_{2})}{(\text{\textgreek{W}}-0)(\text{\textgreek{W}}+\text{\textgreek{W}})}e^{-i[\text{\textgreek{W}}+\frac{1}{2}(\omega_{1}+\frac{\text{\textgreek{W}}_{2}^{2}}{\text{\textgreek{W}}^{2}}\omega_{2})]t}+\frac{\text{\textgreek{W}}_{1}(-\text{\textgreek{W}}-\omega_{2})}{(-\text{\textgreek{W}}-0)(-\text{\textgreek{W}}-\text{\textgreek{W}})}e^{-i[-\text{\textgreek{W}}+\frac{1}{2}(\omega_{1}+\frac{\text{\textgreek{W}}_{2}^{2}}{\text{\textgreek{W}}^{2}}\omega_{2})]t}\nonumber \\
 & =\frac{\text{\textgreek{W}}_{1}}{\text{\textgreek{W}}}e^{-i\frac{1}{2}(\omega_{1}+\frac{\text{\textgreek{W}}_{2}^{2}}{\text{\textgreek{W}}^{2}}\omega_{2})t}\cos\text{\textgreek{W}}t,
\end{align}
\begin{align}
w(t) & \simeq\text{\textgreek{W}}_{1}\text{\textgreek{W}}_{2}\{\frac{e^{-i\frac{\text{\textgreek{W}}_{1}^{2}}{\text{\textgreek{W}}^{2}}\omega_{2}t}}{(0-\omega_{+})(0-\omega_{-})}+\frac{e^{-i[\text{\textgreek{W}}+\frac{1}{2}(\omega_{1}+\frac{\text{\textgreek{W}}_{2}^{2}}{\text{\textgreek{W}}^{2}}\omega_{2})]t}}{(\omega_{+}-0)(\omega_{+}-\omega_{-})}+\frac{e^{-i[-\text{\textgreek{W}}+\frac{1}{2}(\omega_{1}+\frac{\text{\textgreek{W}}_{2}^{2}}{\text{\textgreek{W}}^{2}}\omega_{2})]t}}{(\omega_{-}-0)(\omega_{-}-\omega_{+})}\}\nonumber \\
 & \simeq-\frac{\text{\textgreek{W}}_{1}\text{\textgreek{W}}_{2}}{\text{\textgreek{W}}^{2}}e^{-i\frac{\text{\textgreek{W}}_{1}^{2}}{\text{\textgreek{W}}^{2}}\omega_{2}t}+\frac{\text{\textgreek{W}}_{1}\text{\textgreek{W}}_{2}}{2\text{\textgreek{W}}^{2}}e^{-i[\text{\textgreek{W}}+\frac{1}{2}(\omega_{1}+\frac{\text{\textgreek{W}}_{2}^{2}}{\text{\textgreek{W}}^{2}}\omega_{2})]t}+\frac{\text{\textgreek{W}}_{1}\text{\textgreek{W}}_{2}}{2\text{\textgreek{W}}^{2}}e^{-i[-\text{\textgreek{W}}+\frac{1}{2}(\omega_{1}+\frac{\text{\textgreek{W}}_{2}^{2}}{\text{\textgreek{W}}^{2}}\omega_{2})]t}\nonumber \\
 & \simeq-\frac{\text{\textgreek{W}}_{1}\text{\textgreek{W}}_{2}}{\text{\textgreek{W}}^{2}}[e^{-i\frac{\text{\textgreek{W}}_{1}^{2}}{\text{\textgreek{W}}^{2}}\omega_{2}t}-e^{-i\frac{1}{2}(\omega_{1}+\frac{\text{\textgreek{W}}_{2}^{2}}{\text{\textgreek{W}}^{2}}\omega_{2})t}\cos\text{\textgreek{W}}t].
\end{align}
When $\text{\textgreek{W}}t=\pi$,
\begin{align}
w(t) & =-\frac{\text{\textgreek{W}}_{1}\text{\textgreek{W}}_{2}}{2\text{\textgreek{W}}^{2}}[e^{-i\frac{\text{\textgreek{W}}_{1}^{2}}{\text{\textgreek{W}}^{2}}\frac{\pi}{\Omega}\omega_{2}}+e^{-i\frac{1}{2}(\omega_{1}+\frac{\text{\textgreek{W}}_{2}^{2}}{\text{\textgreek{W}}^{2}}\omega_{2})\frac{\pi}{\Omega}}]\nonumber \\
 & =-\frac{\text{\textgreek{W}}_{1}\text{\textgreek{W}}_{2}}{\text{\textgreek{W}}^{2}}\{e^{-i\frac{\pi\text{\textgreek{W}}_{1}^{2}}{\text{\textgreek{W}}^{3}}(\text{\textgreek{D}}-i\frac{\kappa}{2})}+e^{-i\frac{1}{2}[(\delta+\frac{\text{\textgreek{W}}_{2}^{2}}{\text{\textgreek{W}}^{2}}\text{\textgreek{D}})-i(1+\frac{\text{\textgreek{W}}_{2}^{2}}{\text{\textgreek{W}}^{2}})\frac{\kappa}{2}]\frac{\pi}{\Omega}}\}.
\end{align}
In order to obtain nearly-complete polarization, the two following
conditions should be fulfilled, i.e.\begin{subequations}
\begin{align}
e^{-i\frac{\pi\text{\textgreek{W}}_{1}^{2}}{\text{\textgreek{W}}^{3}}\text{\textgreek{D}}} & =e^{-i\frac{1}{2}(\delta+\frac{\text{\textgreek{W}}_{2}^{2}}{\text{\textgreek{W}}^{2}}\text{\textgreek{D}})\frac{\pi}{\Omega}},\\
1 & \gg e^{-\frac{\pi\text{\textgreek{W}}_{1}^{2}}{2\text{\textgreek{W}}^{3}}\kappa},e^{-\frac{\text{\textgreek{W}}^{2}+\text{\textgreek{W}}_{2}^{2}}{4\text{\textgreek{W}}^{3}}\pi\kappa},
\end{align}
\end{subequations}or equivalently\begin{subequations}
\begin{align}
\delta & =\frac{2\text{\textgreek{W}}_{1}^{2}-\text{\textgreek{W}}_{2}^{2}}{\text{\textgreek{W}}^{2}}\text{\textgreek{D}},\\
\kappa & \ll\frac{\text{\textgreek{W}}^{3}}{\text{\ensuremath{\pi}}\Omega_{1}^{2}},\frac{4\text{\textgreek{W}}^{3}}{\text{\ensuremath{\pi}(\ensuremath{\text{\textgreek{W}}^{2}}+\ensuremath{\text{\textgreek{W}}_{2}^{2}})}}.
\end{align}
\end{subequations}For $\text{\textgreek{W}}_{1}=\text{\textgreek{W}}_{2}$,
we have\begin{subequations}
\begin{align}
\delta & =\frac{1}{2}\text{\textgreek{D}},\\
\text{\textgreek{W}} & \gg\frac{3}{8}\pi\kappa.
\end{align}
\end{subequations}When the above condition is fulfilled, to the first
order of $\kappa$, the polarization deviates from the unity as
\begin{align}
1-\vert w(t)\vert^{2} & \simeq1-\frac{\text{\textgreek{W}}_{1}^{2}\text{\textgreek{W}}_{2}^{2}}{\text{\textgreek{W}}^{4}}\{e^{-\frac{\pi\text{\textgreek{W}}_{1}^{2}}{2\text{\textgreek{W}}^{3}}\kappa}+e^{-(1+\frac{\text{\textgreek{W}}_{2}^{2}}{\text{\textgreek{W}}^{2}})\frac{\pi}{2\Omega}\kappa}+2\mathrm{Re}[e^{-i\frac{\pi\text{\textgreek{W}}_{1}^{2}}{\text{\textgreek{W}}^{3}}(\text{\textgreek{D}}-i\frac{\kappa}{2})}e^{i\frac{1}{2}[(\delta+\frac{\text{\textgreek{W}}_{2}^{2}}{\text{\textgreek{W}}^{2}}\text{\textgreek{D}})+i(1+\frac{\text{\textgreek{W}}_{2}^{2}}{\text{\textgreek{W}}^{2}})\frac{\kappa}{2}]\frac{\pi}{\Omega}}]\}\nonumber \\
 & =1-\frac{\text{\textgreek{W}}_{1}^{2}\text{\textgreek{W}}_{2}^{2}}{\text{\textgreek{W}}^{4}}\{e^{-\frac{\pi\text{\textgreek{W}}_{1}^{2}}{2\text{\textgreek{W}}^{3}}\kappa}+e^{-(1+\frac{\text{\textgreek{W}}_{2}^{2}}{\text{\textgreek{W}}^{2}})\frac{\pi}{2\Omega}\kappa}+2\cos[-\frac{\pi\text{\textgreek{W}}_{1}^{2}}{\text{\textgreek{W}}^{3}}\text{\textgreek{D}}+\frac{1}{2}(\delta+\frac{\text{\textgreek{W}}_{2}^{2}}{\text{\textgreek{W}}^{2}}\text{\textgreek{D}})\frac{\pi}{\Omega}]e^{-\frac{\pi\text{\textgreek{W}}_{1}^{2}}{2\text{\textgreek{W}}^{3}}\kappa}e^{-\frac{1}{4}(1+\frac{\text{\textgreek{W}}_{2}^{2}}{\text{\textgreek{W}}^{2}})\frac{\pi}{\Omega}\kappa}]\}\nonumber \\
 & \simeq1-\frac{\text{\textgreek{W}}_{1}^{2}\text{\textgreek{W}}_{2}^{2}}{\text{\textgreek{W}}^{4}}\{1-\frac{\pi\text{\textgreek{W}}_{1}^{2}}{2\text{\textgreek{W}}^{3}}\kappa+1-(1+\frac{\text{\textgreek{W}}_{2}^{2}}{\text{\textgreek{W}}^{2}})\frac{\pi}{2\Omega}\kappa+2[1-\frac{\pi\text{\textgreek{W}}_{1}^{2}}{2\text{\textgreek{W}}^{3}}\kappa-\frac{1}{2}(1+\frac{\text{\textgreek{W}}_{2}^{2}}{\text{\textgreek{W}}^{2}})\frac{\pi}{2\Omega}\kappa]\}\nonumber \\
 & =1-\frac{\text{\textgreek{W}}_{1}^{2}\text{\textgreek{W}}_{2}^{2}}{\text{\textgreek{W}}^{4}}[4-3\frac{\pi\text{\textgreek{W}}_{1}^{2}}{2\text{\textgreek{W}}^{3}}\kappa-2(1+\frac{\text{\textgreek{W}}_{2}^{2}}{\text{\textgreek{W}}^{2}})\frac{\pi}{2\Omega}\kappa]\nonumber \\
 & =1-(1-\frac{3\pi}{16\text{\textgreek{W}}}\kappa-\frac{3\pi}{16\Omega}\kappa)\nonumber \\
 & =\frac{3\pi}{8\text{\textgreek{W}}}\kappa.
\end{align}

To the zeroth order of $\omega_{j}$'s, the eigen states are\begin{subequations}
\begin{align}
\left\vert E_{1}\right\rangle  & \simeq\frac{1}{N_{i}}\{[(0-\omega_{1})(0-\omega_{2})-\text{\textgreek{W}}_{2}^{2}]\left\vert0,\uparrow\right\rangle +\text{\textgreek{W}}_{1}(0-\omega_{2})\left\vert-,\uparrow\right\rangle +\text{\textgreek{W}}_{1}\text{\textgreek{W}}_{2}\left\vert-,\downarrow\right\rangle \}\nonumber \\
 & \simeq\frac{\text{\textgreek{W}}_{2}}{N_{1}}(-\text{\textgreek{W}}_{2}\left\vert0,\uparrow\right\rangle +\text{\textgreek{W}}_{1}\left\vert-,\downarrow\right\rangle ),\\
\left\vert E_{2}\right\rangle  & \simeq\frac{1}{N_{2}}\{[(\Omega-\omega_{1})(\Omega-\omega_{2})-\text{\textgreek{W}}_{2}^{2}]\left\vert0,\uparrow\right\rangle +\text{\textgreek{W}}_{1}(\Omega-\omega_{2})\left\vert-,\uparrow\right\rangle +\text{\textgreek{W}}_{1}\text{\textgreek{W}}_{2}\left\vert-,\downarrow\right\rangle \}\nonumber \\
 & \simeq\frac{\text{\textgreek{W}}_{1}}{N_{2}}(\text{\textgreek{W}}_{1}\left\vert0,\uparrow\right\rangle +\Omega\left\vert-,\uparrow\right\rangle +\text{\textgreek{W}}_{2}\left\vert-,\downarrow\right\rangle ),\\
\left\vert E_{3}\right\rangle  & \simeq\frac{1}{N_{3}}\{[(-\Omega-\omega_{1})(-\Omega-\omega_{2})-\text{\textgreek{W}}_{2}^{2}]\left\vert0,\uparrow\right\rangle +\text{\textgreek{W}}_{1}(-\Omega-\omega_{2})\left\vert-,\uparrow\right\rangle +\text{\textgreek{W}}_{1}\text{\textgreek{W}}_{2}\left\vert-,\downarrow\right\rangle \}\nonumber \\
 & \simeq\frac{\text{\textgreek{W}}_{1}}{N_{3}}(\text{\textgreek{W}}_{1}\left\vert0,\uparrow\right\rangle -\Omega\left\vert-,\uparrow\right\rangle +\text{\textgreek{W}}_{2}\left\vert-,\downarrow\right\rangle ),
\end{align}
\end{subequations}where the normalization constants are\begin{subequations}
\begin{align}
N_{1}^{2} & =\text{\textgreek{W}}_{2}^{2}\Omega^{2},\\
N_{2}^{2} & =N_{3}^{2}=2\text{\textgreek{W}}_{1}^{2}\Omega^{2}.
\end{align}
\end{subequations}Here $\left\vert E_{1}\right\rangle $ is the dark state,
while the other two are the bright states. Notice that all expanding
coefficients in the bright states are the same except there is a sign
difference in the expanding coefficients of $\left\vert-,\uparrow\right\rangle $.

In the eigen basis, the time evolution of the initial state $\vert\psi(0)\rangle=\vert0,\uparrow\rangle$
is
\begin{align}
\vert\psi(t)\rangle & =\frac{\text{\textgreek{W}}_{2}\Omega e^{-i\frac{\text{\textgreek{W}}_{1}^{2}}{\text{\textgreek{W}}^{2}}\omega_{2}t}}{(0-\Omega)(0+\Omega)}\left\vert E_{1}\right\rangle +\frac{\sqrt{2}\text{\textgreek{W}}_{1}\Omega e^{-i[\text{\textgreek{W}}+\frac{1}{2}(\omega_{1}+\frac{\text{\textgreek{W}}_{2}^{2}}{\text{\textgreek{W}}^{2}}\omega_{2})]t}}{(\Omega-0)(\Omega+\Omega)}\left\vert E_{2}\right\rangle +\frac{\sqrt{2}\text{\textgreek{W}}_{1}\Omega e^{-i[-\text{\textgreek{W}}+\frac{1}{2}(\omega_{1}+\frac{\text{\textgreek{W}}_{2}^{2}}{\text{\textgreek{W}}^{2}}\omega_{2})]t}}{(-\Omega-0)(-\Omega-\Omega)}\left\vert E_{3}\right\rangle \nonumber \\
 & =-\frac{\text{\textgreek{W}}_{2}e^{-i\frac{\text{\textgreek{W}}_{1}^{2}}{\text{\textgreek{W}}^{2}}\omega_{2}t}}{\Omega}\left\vert E_{1}\right\rangle +\frac{\sqrt{2}\text{\textgreek{W}}_{1}e^{-i\frac{1}{2}(\omega_{1}+\frac{\text{\textgreek{W}}_{2}^{2}}{\text{\textgreek{W}}^{2}}\omega_{2})t}}{2\Omega}(e^{-i\text{\textgreek{W}}t}\left\vert E_{2}\right\rangle +e^{i\text{\textgreek{W}}t}\left\vert E_{3}\right\rangle ).
\end{align}
Because there is a sign difference in the expanding coefficients of
$\left\vert-,\uparrow\right\rangle $ in $\left\vert E_{2}\right\rangle $
and $\left\vert E_{3}\right\rangle $, the probability in $\left\vert-,\uparrow\right\rangle $
vanishes as long as $\Omega t=\pi n$ with $n$ being integer. To
summarize, we utilize the dark state and quantum interference to achieve
nearly-complete polarization of the nuclear spin.
\end{widetext}

\end{document}